\documentclass[utf8]{frontiersinFPHY_FAMS} % Vancouver Reference Style (Numbered) for articles in the journals "Frontiers in Physics" and "Frontiers in Applied Mathematics and Statistics" 

\setcitestyle{square} % for articles in the journals "Frontiers in Physics" and "Frontiers in Applied Mathematics and Statistics" 
\usepackage{url,hyperref,lineno,microtype,subcaption}
\usepackage[onehalfspacing]{setspace}
\usepackage{changes}
\def\keyFont{\fontsize{8}{11}\helveticabold }
\def\firstAuthorLast{Maria C. Babiuc Hamilton {et~al.}} %use et al only if is more than 1 author
\def\Authors{Maria C. Babiuc Hamilton\,$^{1,*}$ and Joseph I. Powell\,$^{1,2}$}
\def\Address{$^{1}$Marshall University, Department of Mathematics and Physics, Huntington, WV, United States \\
$^{2}$Undergraduate Student}
\def\corrAuthor{Maria C. Babiuc Hamilton}

\def\corrEmail{babiuc@marshall.edu}

\begin{document}
\onecolumn
\firstpage{1}

%\title {Signatures of Nuclear Isomers in Gamma-Ray Bursts from Binary Neutron Star Mergers} 
\title{How Do Nuclear Isomers Influence the Gamma-Ray Bursts in Binary Neutron Star Mergers?}

\author[\firstAuthorLast ]{\Authors} %This field will be automatically populated
\address{} %This field will be automatically populated
\correspondance{} %This field will be automatically populated

\extraAuth{}% If there are more than 1 corresponding author, comment this line and uncomment the next one.
%\extraAuth{corresponding Author2 \\ Laboratory X2, Institute X2, Department X2, Organization X2, Street X2, City X2 , State XX2 (only USA, Canada and Australia), Zip Code2, X2 Country X2, email2@uni2.edu}

\maketitle

\begin{abstract}

%%% Leave the Abstract empty if your article does not require one, please see the Summary Table for full details.
\section{}
%For full guidelines regarding your manuscript please refer to \href{https://www.frontiersin.org/guidelines/author-guidelines}{Author Guidelines}.
Neutron star mergers are astrophysical `gold mines,' synthesizing over half of the elements heavier than iron through rapid neutron capture nucleosynthesis. The observation of the binary neutron star merger GW170817, detected both in gravitational waves and electromagnetic radiation, marked a breakthrough. One electromagnetic component of this event, the gamma ray burst GRB 170817A, has an unresolved aspect: the characteristics of its prompt gamma-ray emission spectrum. 
In this work, we investigate how gamma-ray spectra in such GRBs may be influenced by de-excitations from isomeric transitions. Our study begins with a review of current knowledge on GRB structure and of r-process nucleosynthesis in neutron star collisions, focusing on the role of nuclear isomers in these settings. We then test our hypothesis by developing criteria to select representative isomers, based on known solar element abundances, for modeling GRB spectral characteristics. We integrate these criteria into an interactive web page, facilitating the construction and analysis of relevant gamma-ray spectra from isomeric transitions. Our analysis reveals that three isomers—$_{90}\text{Zr}$, $_{207}\text{Pb}$, and $_{89}\text{Y}$—stand out for their potential to impact the prompt GRB spectrum due to their specific properties.
This information allows us to incorporate nuclear isomer data into astrophysical simulations and calculate isomeric abundances generated by astrophysical r-processes in neutron star mergers and their imprint on the detected signal.
%As a primary goal, the abstract should render the general significance and conceptual advance of the work clearly accessible to a broad readership. References should not be cited in the abstract. Leave the Abstract empty if your article does not require one, please see the Article Types on every Frontiers journal page for full details

\tiny
 \keyFont{ \section{Keywords:} binary neutron mergers, gamma-ray bursts, rapid neutron capture nucleosynthesis, nuclear isomers, isomeric transition} %All article types: you may provide up to 8 keywords; at least 5 are mandatory.
\end{abstract}

\section{Introduction}
%the introduction should be succinct, with no subheadings 
%For Original Research Articles \citep{conference}, Clinical Trial Articles \citep{article}, and Technology Reports \citep{patent}, \citep{book}. For Case Reports the Introduction should include symptoms at presentation \citep{chapter}, physical exams and lab results \citep{dataset}.

%For Original Research Articles \cite{conference}, Clinical Trial Articles \cite{article}, and Technology Reports \cite{patent}, \cite{book}. For Case Reports the Introduction should include symptoms at presentation \cite{chapter}, physical exams and lab results \cite{dataset}.

%Background: 
%Rationale: 
%Statement of purpose: 
%Summary

The collision of two neutron stars injects massive amounts of matter at high energy into the surrounding environment and release an enormous amount of energy that can be detected as gravitational waves and light, emitted across the electromagnetic spectrum on various timescales \cite{arXiv:2005.14135}. 

The first detection of gravitational waves (GW) from a binary neutron star (BNS) collision, named GW170817 \cite{arXiv:1710.05832}, was accompanied by a weak, off-axis $\gamma$-ray burst (GRB 170817A) \cite{arXiv:1710.05834}, matter outflow \cite{arXiv:1705.02598} and an optical-to-infrared transient (AT2017gfo), called a kilonova \cite{arXiv:1710.05843}, that emitted a broad array of electromagnetic (EM) radiation \cite{arXiv:1803.07595}, in accord with the predictions for radioactive decay of elements. 
Analysis of the data gathered from this collision provided us with a wealth of new insights into many yet unknown aspects of a neutron star \cite{arXiv:1102.5735}, such as its internal structure and radius \cite{arXiv:1805.11581}, 
%arXiv:1805.11963, arXiv:1807.06437, 
the amplification of the magnetic fields during merger \cite{arXiv:2004.10105}, and the outflow of matter during the collision process \cite{arXiv:1809.06843}.
Further theoretical studies of this twin detection in GW and EM radiation allowed scientists to identify for the first time an element heavier than iron, namely strontium \cite{arXiv:1910.10510}, thus confirming BNS mergers as sites where heavy elements are formed \cite{arXiv:1801.01141}. 
These discoveries established neutron star mergers as veritable “gold mines,” because more than half of all the elements heavier than iron can be forged in the collision \cite{arXiv:2008.04333} through the rapid-process (r-process) nucleosynthesis, the dominant mechanism through which neutrons are promptly captured by seed nuclei to build up heavier elements before they radioactively decay \cite{arXiv:1901.01410}.

The detection of the accompanied GRB 170817A signal \cite{arXiv:1710.05446, arXiv:1710.05449} also unambiguously confirmed compact binary mergers as sources of gamma-ray bursts (GRB) \cite{arXiv:1808.04831, arXiv:1905.02665}.
However, one of the puzzling questions that still remains is the mechanism of the highly collimated short GRB  \cite{arXiv:1711.00243, arXiv:1807.04756, arXiv:1808.05794}. 
 %arXiv:1710.05857, arXiv:1710.05897, 
During the collision, a large amount of highly relativistic matter is projected outwards, along the axis of rotation. 
These particles are trapped, collimated and accelerated to near-light speed by strong magnetic fields amplified during the merger, and are carried to long distances, spanning entire galaxies. 
In their travel, these particles might convert kinetic energy into $\gamma$-ray (EM) energy through synchrotron emission, inverse Compton processes and dissipation due to internal shocks. 
Although great strides have been made in understanding the nature of GRB emission, one open issue remaining is the interpretation of the radiative mechanisms responsible for the prompt GRB spectrum.
At the beginning, the $\gamma$-ray energies have a different origin than the synchrotron radiation, suggesting that a more complex model is required to fit the data. 
%One aspect that remains unexplained is the no-thermal $\gamma$-ray spectrum of the prompt emission. 
This emission might contain as well $\gamma$-ray photons from nuclear decay of the heavy elements produced in the ejecta \cite{arXiv:1406.4440}.
%, although their signal will overlap with the synchrotron, inverse Compton and thermal emission. 
The heavy nuclei produced in this outflow of matter by the r-process nucleosynthesis are stripped of electrons, and many of them are unstable.
While still in excited nuclear states, some of these nuclei can be trapped by the strong magnetic field and beamed to high velocities.  
The radioactive decay of these unstable isotopes will release nuclear energy in form of $\gamma$-ray, neutrino and particle products in the jet. 
While the neutrino will escape freely, the electrons will annihilate with the positrons, and the other particles will be thermalized through collisions, Coulomb forces and inverse Compton scattering, producing additional $\gamma$-rays.

A portion of these newly formed, highly excited heavy isotopes will be in isomeric, metastable nuclear states with lifetimes long enough to enable them to be distinguished from other nuclear states. 
The nuclear isomers that retain their metastable characteristics in highly energetic astrophysical environments such as BNS mergers are called astromers \cite{arXiv:2010.15238} and may play a significant role in determining not only the abundance of the elements in the universe \cite{arXiv:2011.11889}, but also the spectral appearance of GRBs from such events.
Scientists have started to consider the influence of astromers in the r-process nucleosynthesis in connection with the light-curve of the EM signal from the ejecta of BNS mergers \cite{arXiv:1803.08335, arXiv:2001.10668}. 
However, this aspect is still in its infancy, and none has explored their connection to GRB $\gamma$-ray emission. 

One aspect that remains unexplained is the non-thermal $\gamma$-ray spectrum of the prompt emission.
Our research aims to bridge this gap by investigating whether $\gamma$-ray de-excitations from isomeric transitions of excited nuclei created in the r-process nucleosynthesis during BNS mergers can contribute to the spectrum of the GRBs associated to these events.
An accurate understanding of this $\gamma$-ray emission through isomeric transition is challenging because we have limited knowledge of the nuclear physics that operates the r-process nucleosynthesis of heavy elements during a BNS collision in the presence of isomers. 
Similar to the stable elements, isomers are thought to conglomerate around the `magic numbers' of neutron shell closure and can lead to a change in the elemental abundances produced in nucleosynthesis, or even influence the path of the nuclear reactions \cite{arXiv:2103.09392}. 
%Moreover, in the extremely energetic environment of the merger, nuclei in isomeric states are thought to have a significantly smaller effective lifetime due to thermally mediated transitions between nucleons \cite{arXiv:2103.09392}.
Our endeavor to connect isomeric $\gamma$-ray de-excitations with the observed GRB spectrum is feasible, although ambitious. 
To calculate the emission from such transitions, we must know the species of nuclides inside the merger ejecta, and their abundances. 
Moreover, the fate of the $\gamma$-ray photons generated through isomeric decay will be sensitive to the thickness of the ejecta, and will be affected by superpositions with the $\gamma$-rays generated through fission, annihilation radiation, synchrotron emission and thermalization, as well as by the line broadening caused by the Doppler effect during the relativistic expansion of the jet. 

Our paper is organized as follows: we begin with a discussion on GRB properties, and continue with introducing known facts about the r-process nucleosynthesis in BNS collisions, as well as the concept of isomeric transition. 
We continue with presenting our compilation of the relevant isomeric elements likely to be created during the r-process, from \cite{arXiv:2208.01028}. 
We explain our criterion of selecting representative isomers, accounting for the typical time of the r-process, the time necessary for the jet to form and break through the ejecta surrounding the merger, and the time of the GRB burst. 
We carefully cross-correlate the identified isomers with the cosmic abundance of elements observed in the solar system \cite{arXiv:1912.00844} and estimate the number of isomers created relative to the number of baryons in the ejecta. 
We implement our selection criteria in a Python program embedded into an interactive web page, accessible as an open-source shareable data application.
We analyze this data, constructing a set of relevant $\gamma$-ray spectra and light-curves associated with the most promising isomers. 
In a later work, we will refine our model and will compare our results with GRB170817A as well as with other GRBs thought to have originated from double neutron star mergers.

Our new connection between the GRB emission from known astrophysical sites of r-process nucleosynthesis such as BNS mergers and the $\gamma$-ray signature of isomeric transition might be a step forward towards explaining spectral features in past and future GRB detections from compact sources.
The information we provide can be also incorporated into detailed astrophysical simulations, in order to calculate with accuracy the characteristic isomeric abundance produced by BNS collisions and to generate light curves that can be then validated by comparison with detected GRB data.
This is just the beginning and much remains to be learned about the impact of isomers on the creation of heavy elements in astrophysical nucleosynthesis r-processes and on the physics of GRB. 
Our study could contributes to the elucidation of the intriguing mechanism behind the spikes in the spectrum of the GRB.  
Although besides strontium, there is currently no reported detection of elements created during BNS mergers in the light curves from GRB 170817A \cite{arXiv:1710.05449}, future combined GW/GRB detections, with more sensitive instruments, will occur. 
These detections will be needed to elucidate the mechanisms behind $\gamma$-ray emission in the GRB spectrum, and distinguish the role played by isomeric transition, thus validating or disproving our hypothesis.      

\section{GRB Structure}

%%%%%%%%%%%
%%================================%%
%{\bf TODO: -- add ...}
%%================================%%

%%%%%%%%%%%%%%%%%%%%%%%%%%%%%%%%%%%%%%
Gamma ray bursts (GRBs) have been a focal point of research since their initial observation in 1973 by the Vela satellite \cite{gammaray}. 
It is agreed that their $\gamma$-ray emission follows roughly the same pattern, starting with a short, spectrally hard burst, followed by a longer tail of spectrally softer emission, and ending with a long-lasting multi-wavelength afterglow \cite{arXiv:astro-ph/9807272, arXiv:2201.09796}.
Despite the detection of over $12,000$ GRBs since their discovery, and extensive research on this topic, the jet formation mechanism remains elusive \cite{arXiv:2308.04485, arXiv:2308.12709}.

Based on the observed time frame of the $\gamma$-ray emission, astronomers have categorized GRBs in two groups: long ($>2\mathrm{s}$) and short ($<2\mathrm{s}$) bursts. The short-duration GRBs are considered to originate when two compact objects merge, while long GRBs could result from a collapsing massive star, or supernova. 
However, recent discoveries of long-duration bursts such as GRB 211211A \cite{arXiv:2205.08566, arXiv:2204.10864} and GRB 230307A \cite{arXiv:2312.01074} show evidence that they are consistent with the detection of an associated kilonova.
Their spectra present extreme variability, flares and quasi-periodic substructure, characteristic to the formation of a neutron star remnant prior to the final collapse \cite{arXiv:2305.12262, arXiv:2303.08062, arXiv:2301.02864}.
These discoveries point towards a new class of long GRBs originating from mergers of neutron stars. 

The mainstream explanation of the GRB engine is centered on the Blandford-Znajek mechanism. This theory requires the existence of a black hole (BH) surrounded by an accretion disk. The accretion disk supports large-scale aligned magnetic fields, which thread through the central black hole.
This magnetic field extracts spin energy from the black hole, directing it into a low-mass jet, and accelerating it to relativistic speeds \cite{blandford}.
Numerical simulations do report that during the collision of two neutron stars, the jet is formed after the remnant collapses to a black hole \cite{arXiv:2104.12410}.
An alternative explanation for the GRB engine is a fast-spinning, strongly magnetized neutron star, called a magnetar, that can dump its rotational energy into the Poynting flux, who transports the energy of the magnetic fields in form of electromagnetic radiation from the star to the jet. This mechanism could accelerate a small amount of matter to very high energies, producing the relativistic jet. 
Indeed, other numerical simulations prove that magnetars formed as remnants of BNS mergers are viable engines able to launch GRBs and power kilonovae \cite{arXiv:1905.01190, arXiv:2003.06043}.

The current understanding of GRB production involves a compact star (either a magnetar or a black hole with an accretion disk) generating a large amount of highly relativistic particles. These particles extract energy from the compact object through the Poynting flux, carrying them over large distances.
During their travel, the stream of particles become optically thin and might experience shocks, and convert their kinetic energy into internal energy. The observed $\gamma$-ray\added{s} are subsequently emitted through synchrotron radiation or inverse Compton processes when relativistic electrons are being accelerated in magnetic fields \cite{arXiv:1710.05823, arXiv:1905.01190}. 

The GRB 170817A jet was detected at $t_0=1.75\mathrm{s}$ after the peak in the GW signal, and lasted around $t_j = 2 \mathrm{s}$, starting with an initial spike in $\gamma$-ray energy of about $0.5 \mathrm{s}$, followed by a broader and less intense tail \cite{arXiv:1710.05823}. 
Its $\gamma$-ray emission was less luminous than known short-duration GRBs, leading scientists to infer that the emission was off-axis \cite{arXiv:1710.05869}, and subsequently scattered while passing through the merger ejecta, with a peak in photon energy of about $\textrm{5 MeV}$ and a narrow half-opening of $\approx 2.1^{+2.4}_{-1.4}$ degrees, viewed at an angle of $23^{+5}_{-3}$ degrees \cite{arXiv:2306.15488, arXiv:2305.06275, arXiv:2306.16795}. 
This model classified GRB 170817A as a typical short GRB, favoring a quasi-universal jet structure, with the differences being caused by extrinsic factors, such as density of the particles in the jet, viewing angle or interaction with the surrounding medium \cite{arXiv:1905.01190}. 

Numerical simulations showed that the time delay between the merger and the start of the jet was due to (1) the time necessary for engine activation and (2) the time for the jet to break out of the surrounding environment \cite{arXiv:2104.12410}. 
After a careful examination, it was shown that the time delay should include three terms, namely (1) the time to launch the jet, (2) the time for the jet to break out from the surrounding medium and (3) the time to reach the emission site. 
The fact that the time delay for GRB 170817A correlates with the pulse duration was interpreted to indicate that the time delay is dominated by the duration for the jet to travel to the emission radius, estimated to be at large distance ($\approx 10^{15} \textrm{cm}$) from the progenitor \cite{arXiv:1905.00781}.
As consequence, the time delay cannot be used to diagnose the jet launching mechanism. %, for a Poynting flux dominated outflow. 

In this model, the $\gamma$-rays are produced far away from the engine, in the circumstellar region populated by the outflow of gas ejected during the merger, driven by magnetic fields. 
The radiation was released from a broad outflow with a large opening angle, and subsequently collimated, partly by the large-scale, ordered magnetic field, and partly due to the ultra-relativistic motion of the particles in the jet \cite{arXiv:2306.16795}. 
Relativistic outflows are strongly beamed, such that the observer sees only the beaming angle, proportional to $1/\Gamma$, where $\Gamma$ is the Lorentz factor, a measure of the relativistic effects experienced by objects moving close to the speed of light. 
The estimated values for the Lorentz factor of the bulk matter in the jet are very large, between $100$ and $1000$ \cite{arXiv:2312.02259}.
Particles accelerated to such relativistic speeds posses extremely high energy and emit synchrotron radiation in strong magnetic fields. 

Nonetheless, many GRB spectra deviate from the expectations of this synchrotron emission.
For example, the light-curves of the prompt emission are irregular. One of the hypothesis is that this variability is due to internal shocks \cite{arXiv:1711.00243, arXiv:1905.01190, arXiv:2009.01773}.
Relativistic jets can generate shock waves because the inner engine produces inhomogeneities, and shells of particles in the jet travel at different velocities. 
However, the internal shock model of GRBs is inefficient in converting the kinetic energy of the particles in the jet into $\gamma$-ray radiation, known as the `low-efficiency problem.' 
This was replaced with a model in which a `fireball,' moving at a relativistic speed, is launched by a fast-rotating black hole or magnetar. In this case, the internal shocks are supplemented with the external shock mechanism \cite{arXiv:astro-ph/9810256, arXiv:1606.00311, arXiv:2306.16795}. 
Because the velocity of the particles in the jet is larger than the speed of sound, the beamed ejecta will form a cocoon when it plows through the surrounding medium and this interaction modulates the synchrotron radiation \cite{arXiv:1710.05896}. 
The jet and cocoon combination forms a `structured' jet, which avoids the underlying mechanism. 
A structured jet has a uniform, ultra-relativistic core, surrounding by a power-law decaying wind, forming a two-jet component, with a relativistic core and a slightly slower outer jet \cite{arXiv:2206.11088}. 
To explain photon energies greater than $10\textrm{GeV}$, this double-jet picture is modified into a narrow, off-axis jet with a high Lorentz factor, and a wide, on axis jet with a small Lorentz factor, the interaction between them forming reverse shock regions that could explain the $\mathrm{GeV}$ flares observed in some GRBs \cite{arXiv:2311.01705}.
%, arXiv:2311.02859, arXiv:2311.01710}.  

This way, a unified picture of both short and long GRBs from compact binary mergers emerges, based on a structured jet launched by a common central engine, which avoids the underlying mechanism \cite{arXiv:1710.05436, arXiv:2309.00038}.
The peak luminosity distribution of the long and short GRBs could be also fitted to a triple power law, implying that both types of GRBs could be produced by the same mechanism \cite{arXiv:2306.03365, arXiv:2311.02859}. 
But if indeed the $\gamma$-ray emission took place in an optically thin region, far away from the central engine, the shock emission components are suppressed, and some other mechanisms may be at play \cite{arXiv:2312.02259}.
For example, recent simulations show that before the emergence on the jet from the neutron star remnant formed after the BNS merger, a UV/blue precursor signal is generated, that can `seed' the ultra relativistic GRB jet \cite{arXiv:2303.12284}. 

Spectral data alone might not be enough to discern between various models and to asses their viability. 
Polarization measurements of the GRB prompt emission, in principle, have the potential to address many of these questions. 
However, such measurements have only been obtained for a limited number of bursts and thus have limited statistical significance \cite{arXiv:2306.16634}.
Looking ahead, joint detections of GW/GRB events, coupled with polarization data from the accompanying GRB, will help understand these astrophysical phenomena.
%In the future, joint detections of GW/GRB with polarization data from the accompanied GRB will become possible.
%Looking ahead, the prospect of joint detections of GW/GRB events, coupled with polarization data from the accompanying GRB, will become possible.
%will open new avenues for understanding these complex astrophysical phenomena

%%%%%%%%%%%%%%%%%%%%%%%%%%%%%%%%%%%%%
	
\section{r-process nucleosynthesis and Isomers}
%%%%%%%%%%%%%%%%%%%%
The r-process consists of a series of reactions in which nuclei capture neutrons rapidly, leading to the creation of heavy elements. This process is believed to occur at high temperatures, in extremely neutron-rich environments.
While the r-process was long associated with supernovae, recent studies indicate that BNS mergers, with their more neutron-rich environments, are likely the predominant sites for heavy r-process element production \cite{arXiv:1406.2687, arXiv:2109.09162, arXiv:2208.14026}.
More than half of the heavy elements found in nature are produced through the r-process, some elements forming exclusively or almost so by this mechanism \cite{arXiv:1901.01410}. 
The detailed pathways of producing these heavy elements are still unsettled \cite{arXiv:2305.03664} and the lack of confidence in the neutron capture rate predictions makes the calculation of final abundances in the r-process difficult \cite{arXiv:1906.05002}. 

The r-process operates in two distinct phases: an initial period in which neutron captures dominate, and a subsequent state characterized by $\beta$-decay, leading to the creation of new elements with increasingly heavier proton numbers.
The timescale for neutron capture is significantly faster than that of $\beta$-decay. Neutrons are absorbed rapidly until a statistical equilibrium is reached, a point known as `neutron drip line,' where neutron separation energy becomes zero.
Here, a neutron shell closure is reached, known as `freeze-out,' where rapid neutron capture ceases \cite{arXiv:1406.2687}. 
This occurs when the neutron-rich environment becomes depleted, and the neutron capture rate drops significantly. At this stage, the nucleus is no longer able to capture neutrons effectively, marking the end of the rapid neutron capture phase and leading to a shift towards $\beta$-decay, where neutrons in the nucleus transform into protons, creating new elements.
These nuclei, formed post freeze-out, act as seeds for subsequent r-processes, continuing to capture neutrons and forming increasingly heavy nuclei.
`Kinks' in the r-process occur at neutron number shell closures, specifically around $N=50, 82,$ and $126$. At these points, nuclei are more stable and resist further neutron capture, leading to an accumulation of material. These kinks influence the distribution of elements produced in the r-process, leading to observable patterns in the abundance of elements.
The r-process culminates at the `magic number' $N=184$, that marks the third r-process peak, signaling the completion of neutron capture and $\beta$-decay.
This happens at about $t_r=1.34\text{s}$ since the beginning of the process, when unstable elements with large atomic mass, $A\approx 240$, are created \cite{arXiv:1411.0974}. 
This instability leads to fission, where the heavy nuclei split into smaller ones, typically in the $A<140$ range, releasing additional neutrons and a significant amount of energy, detected as observable electromagnetic emission associated with neutron star mergers.

The r-process produces a variety of heavy elements, in agreement with the solar system abundance.
Within the energetic collision environment of BNS mergers, a diverse range of conditions leads to various nuclear nucleosynthesis products. Extremely high densities favor the formation of heavy nuclei, while high temperatures tend to produce lighter nuclei. 
%Intermediate conditions typically yield tightly bound nuclei in the atomic mass range of $A=50-60$, such as those in the iron group.
During the coalescence phase, matter is dynamically ejected due to angular momentum conservation within milliseconds, moving at mildly relativistic speeds. This tidal ejecta, dense and moderately heated, retains its original low electron fraction, facilitating the `strong' r-process nucleosynthesis leading to heavier elements \cite{arXiv:1406.2687}.
The production of heavy elements in this ejecta competes with its rapid expansion, that reduces the neutron density and temperature.
During collision, temperatures reach values high enough to dissociate nuclei into free nucleons, and neutrinos become the primary cooling mechanism. At this point, amplified magnetic fields and neutrino winds eject neutron-rich material along the rotation axis, potentially enhancing the production of heavy elements that can be trapped in the jet \cite{arXiv:1405.6730, arXiv:1908.02350}.
The remnant formed after the merger acquires a neutron rich accretion disk, heated to high temperature by friction and irradiation with neutrinos, favoring the continuation of the r-process \cite{arXiv:2112.00772}.

Considering all components – dynamic ejecta, neutrino winds, and outflows from accretion disks – compact binary mergers produce the heaviest r-process nuclei, contributing significantly to the solar r-process abundance \cite{arXiv:2205.05557}.
The early dynamic ejecta, emerging from the spiral arms, stay very neutron rich and lead to strong r-processes, while the late ejecta will produce weaker r-processes.  
Simulations suggest that the mass ratio of the binary affects the range of elements produced, leading to variations in the r-process products across different events.
 \cite{arXiv:2206.02273, arXiv:2305.03664}.
If the magnetic field is amplified to large values, it will drive winds toward the disk, enhancing the production of heavier elements. 
These studies also show that the outflow from the remnants can produce a blue kilonova, indicating the presence of heavy elements \cite{arXiv:2305.07738}. 
Observations of kilonovae suggest also that the amount and distribution of r-process products can differ from event to event \cite{arXiv:1710.05442}.
As astrophysical models of compact binary mergers become more sophisticated and our understanding of neutron-rich nuclei improves, we move closer to accurately predicting the variety and abundance of heavy elements produced in these cosmic events, shedding light on their contribution to the universal abundance of elements. \cite{arXiv:2007.04442, arXiv:2109.09162, arXiv:2206.02273, arXiv:2208.14026, arXiv:2211.04964}. 

Predictions indicate that $\gamma$-ray emissions from neutron star mergers might include photons from the radioactive decay of heavy isotopes produced in the r-process \cite{arXiv:1808.09833, arXiv:2308.00633}.
Those isotopes can find their way in the GRB jets, carried by neutrino cooling winds and by the magnetic field \cite{arXiv:astro-ph/0405510, arXiv:1406.4440, arXiv:1907.00809}.
They can power the $\gamma$-ray bursts and extend the plateau of their $\gamma$-energy emission \cite{arXiv:2104.04708}.
Direct measurements of these photons could potentially serve as a probe for the r-process nucleosynthesis \cite{arXiv:2205.05407, arXiv:1905.05089}.
This must be supplemented with more robust knowledge of the properties of exotic, neutron-rich nuclei to reduce present nuclear uncertainties, that make it difficult to definitely measure the distribution of heavy isotopes \cite{arXiv:nucl-th/0411081}. 

The GW170817/GRB 170817A/AT2017gfo event was identified as a site of r-process nucleosynthesis, observed in the kilonova's electromagnetic spectrum \cite{arXiv:1801.01141, arXiv:2103.15284}.
Although the r-process nucleosynthesis was confirmed by the observations of this event, no trace of individual elements has been identified, except for strontium \cite{arXiv:1910.10510}.
The high density of the spectroscopic lines of the photons expected to be emitted during the r-process, together with the large velocity of the material ejected during the collision produces line blending and smooths the spectra.
This uncertainty complicates the accurate quantification of the heavy element abundances associated with a GRB \cite{arXiv:2308.00633}.
Nuclear data is essential to predict the specific elements that are created in an observed astrophysical environment, and to connect observed abundances and kilonova features to astrophysical conditions and constrains on the nucleosynthesis sites \cite{arXiv:2205.05407}.

Isomers, which are metastable excited states of the same atomic nucleus, can significantly influence $\gamma$-ray emission, and not accounting for them may lead to underestimations of the emission \cite{arXiv:2107.02982}.
If the corresponding lifetimes are of the same order of magnitude as the timescales of the environment, isomers must be treated explicitly \cite{arXiv:1803.08335}.
In the energetic environment of the collision, isomers may either accelerate their decay, slow it down and act as energy storage, or remain unaffected in their half-life. \cite{arXiv:2010.15238, arXiv:2011.11889, arXiv:2103.09392}.
Particularly, isomers could contribute to the early, blue component of kilonova emissions, as observed in GW170817 \cite{arXiv:2001.10668}, potentially allowing outflow towards heavier masses via isomeric branches \cite{arXiv:nucl-th/0411081}.
Near the magic numbers $A \approx 80, 130, 195$ marking the `waiting point' where the r-process temporarily slows down, the excitation energy and the number of isomers increase \cite{arXiv:2208.01028}.
As a result, these points accumulate a higher concentration of nuclei, including isomeric states, that become preferentially populated at these three main peaks of the r-process \cite{arXiv:nucl-th/0411081, arXiv:0803.1700}. 
An important question to answer is how do isotopes reach the isomeric excited state, because promoting nucleons to excited states is hindered due to nuclear recoil (Mossbauer effect, \cite{isomers}). To achieve excitation, the energy of the $\gamma$-ray photon must exceed the transition band energy.
Isomer activation can occur either through the capture of higher-energy $\gamma$-ray photons or via nuclear excitation through thermal excitation at high temperatures \cite{arXiv:2312.09129, arXiv:2401.05598}.
Moreover, when nuclei move at relativistic speeds, they can reach isomeric states by absorbing radiation in ultraviolet and X-rays \cite{arXiv:2106.06584}. 
Internal conversion, involving the ejection of an inner orbital electron, competes with $\gamma$-ray emission, unless the nuclei are in a completely ionized state, a condition found in the atmosphere of the merger \cite{arXiv:2010.15238}.

Understanding these intricate nuclear processes in the astrophysical environment of a neutron star merger and their imprint in the emitted EM radiation from BNS mergers not only sheds light on the complex mechanisms of r-process nucleosynthesis but could also enhance our knowledge of GRB's $\gamma$-ray emission.

\section{Methods}

%%%%%%%%%%%%%%%%%%
%1-3 pages double spaced
%This is the most important part of the paper, and where you need to go into details. Start by briefly re-stating the purpose and approach of your study

%Describe the design:
%Name your study design, and all the instrumentation and software you employed.
%Describe Strategy/Procedures:
%Outline the major steps involved.
%Consider providing a graph.
%Describe Analyses Employed:
%Describe analyses of data, in as much detail as the space allows.
%Take care to spell out the analytical steps in more detail.
%Provide citations supporting your methodological and analytic choices.

Although uncertainties persist in the r-process calculations and their dependency on the astrophysical environment, there is a general agreement that it occurs within a few seconds. Reference \cite{arXiv:1411.0974} indicates that at $t=1.34\text{s}$, the timescale for neutron capture exceeds that of $\beta$-decay, marking the end of the r-process.
The $\gamma$-rays emitted following the r-process are initially trapped within the ejecta and can only be detected after they successfully diffuse through it \cite{barnes}. 
Most of these photons transfer heat and become thermalized, losing their characteristic spectral information.
However, the similar timing observed between the completion of the r-process ($t_r \approx 1.34\text{s}$ ) and the delay in the $\gamma$-ray burst GRB 170817A ($t_0=1.75 \text{s}$) offers a compelling suggestion: the $\gamma$-ray emission from this event may include photons from the de-excitation of heavy elements formed via r-process nucleosynthesis, from parts of the ejecta exposed to the jet funnel. 

It has been previously suggested that emissions from binary neutron star mergers may include gamma rays from nuclear decay \cite{arXiv:2008.03335}.
Additionally, it has been shown that $\beta$-decaying isomer states are more strongly populated than the ground states in stellar environments \cite{arXiv:1804.05657}.
%However, a portion of the ejecta exposed to the jet funnel may still contribute $\gamma$-rays from radioactive decay to the GRB \cite{arXiv:2303.12284}. 
%While the role of isomeric deexcitation in the later stages of a kilonova explosion has been explored \cite{arXiv:2001.10668}, the potential for these $\gamma$-rays to influence the prompt emission of the GRB has not yet been considered.
We propose that a large fraction of these heavy elements is excited into isomeric states and subsequently ejected within the magnetically and neutrino driven wind outflows from the jet engine, contributing to the collimated jet. 
This de-excitation process is likely to contribute to the $\gamma$-ray spectrum observed in GRBs. 
Thus, we put forth the hypothesis that a BNS collision serves as an efficient $\gamma$-ray `factory,' where the primary `raw materials' are the heavy isotopes in their isomeric states, synthesized through r-process nucleosynthesis within the highly energetic and neutron-rich environment of the ejecta.
Their presence may be observed in the $\gamma$-ray emissions through distinct multipolarity signatures, influenced by the spin angular momentum carried by the radiation.

To test this hypothesis, we used the Atlas of Nuclear Isomers \cite{arXiv:2208.01028}, a comprehensive database of experimental data for all known isomers to date. 
This resource includes known properties of each isomer, such as excitation energies, half-lives, decay modes, spins and parities, and energies and multipolarities of isomeric transitions. 
In our analysis, we processed over 2500 isomers and identified which isotopes produced in the r-process are likely to have significant isomeric states that are relevant for $\gamma$-ray production, based on the following two initial criteria:
 \begin{itemize}[nosep]
 \item We start by converting the digital database for the known isomers \cite{arXiv:2208.01028} from its original format into an Excel file to enable compatibility with Python. 
\item We focus only on heavy elements produced in r-process nucleosynthesis and limit our choice to isotopes heavier than iron, starting with an atomic mass number greater than 56.
 \item From these isomers, we select only those that decay via $100\%$ isomeric transitions, because such decays are most relevant for $\gamma$-ray emission and do not alter the chemical properties.
 \end{itemize}

Our next task was to determine the initial quantity of individual isotopes for each element produced and to correlate this value with the respective isomeric form. To achieve this, we relied on estimations of the various types of ejecta present in a neutron star collision \cite{arXiv:1406.2687,  arXiv:astro-ph/0405510}:
\begin{itemize}[nosep]
 \item The early dynamic ejecta emerging from the tidal interaction of the merging neutron stars, typically ranges between $10^{-4} M_{\odot}$ and $10^{-2} M_{\odot}$, depending on the mass ratio and composition \cite{arXiv:1612.03665, arXiv:1710.05836}.
 \item The neutron-rich mass ejected along the rotational axis by the magnetized wind from the merger remnant, is estimated to about $3.5 \times 10^{-3} M_{\odot}$, based on an outflow mass rate of $0.1 M_{\odot}/s$ \cite{arXiv:1405.6730, arXiv:2008.04333}.
 \item The post-merger ejecta surrounding the remnant is calculated to be around  $0.1 M_{\odot}$ \cite{arXiv:1908.02350, arXiv:2205.05557}. % and $0.5 M_{\odot}$.
\end{itemize}

To approximate the mass contributing to isomeric decay, we focused on the mass of the magnetized wind. Additionally, we incorporated a proportion of the mass from both the early dynamic ejecta and the post-merger outflow, that aligns with the jet. 
The exact angular distribution of the ejecta within the jet remains uncertain, varying with the parameters of the merger. However, the multi-wavelength afterglow of GW170817 suggests a stratified geometry of the ejecta, as indicated in \cite{arXiv:1712.03237}. Models range from a top-hat to isotropic fireball geometry, but evidence increasingly supports a structured composition \cite{arXiv:1710.05436, arXiv:1710.05869, arXiv:1712.03237}. The post-merger ejecta, which constitutes the majority of the material, is propelled primarily by the shock at the contact surface during the merger of the two neutron stars. Due to the accumulation of tidal ejecta near the equatorial plane of the binary, this outflow is predominantly directed along the polar axis \cite{arXiv:1711.03112}.
Considering an isotropic ejecta as a starting point, with matter uniformly distributed around the merger site, the mass encompassed within the half-opening angle $\theta_0$ corresponds to the fraction of the spherical surface area covered by the jet $A_j = \Omega r_0^2$ , where $\Omega = 2\pi \left(1 - \cos \theta_0 \right)$ is the solid angle of the jet's cone and $r_0$ is the radius from the central engine at which the jet forms. 
Thus, the effective mass in the jet is: 
\begin{equation}
M_{\text{eff}} = M_{\text{ejecta}} \left (1 - \cos \theta_0 \right),
\label{eq:Meff}
\end{equation}
where $M_{\text{ejecta}}$ is the ejecta mass. This formula takes into account that the jets emanate from both poles.

After calculating the quantity of material expected to influence $\gamma$-ray production in the jet, our next step is to calculate the initial number of isotopes. This involves the following steps:
\begin{itemize}[nosep]
\item We rely on the atomic abundances and the mass fractions for the specific isotopic composition in our solar system, as detailed in Table 9 of \cite{arXiv:1912.00844}, 
We start with the heavy elements that begin forming near the first peak of the r-process, around atomic weight 
$A = 80$, namely with strontium. % \cite{arXiv:2312.01074}.
\item We estimate the fraction of each element produced in merging neutron stars (see Table \ref{tab1}), based on Table S1 in the supplementary material from \cite{johnson}. A revised estimation is found in \cite{busso}. 
\item We calculate the quantity of baryons for each element, rescaled to the abundance of heavy elements baryons, assuming that neutron star collision produce mainly these heavy elements.
\item We rescale this quantity back to only $10\%$ of the matter in the jet assumed to be converted into heavy elements through the r-process \cite{arXiv:1710.05836}. 
\item From the calculated number of baryons for each element produced via the r-process in the jet, we determine the number of atoms and the corresponding mass abundance, in terms of solar mass.
\item We normalize the number of atoms for each element to the total number of atoms in the jet. This normalization eliminates the dependence on the effective mass, and will allow to simply scale the contribution of each isotope to the $\gamma$-ray production in the jet.
\item Lastly, we select among these heavy elements the atoms that admit the previously identified isomeric states, to obtain the type and number of isomers relevant for $\gamma$-ray production.
\end{itemize}
%by calculating the number of nickel (Ni) atoms in the ejecta. We transform in grams the mass of the effective ejecta $M_{\text{eff}}$, divide it by Ni's atomic mass $A=58.7\text{grams/mole}$ and then multiply the result by Avogadro's number. Through this process, we find the total number of Ni atoms, $N_{\text{Ni}}$, to be approximately $2.55 \times 10^{53}$ atoms.
%\item Using the solar elemental abundance data \cite{arXiv:1912.00844}, we rescale the abundance of all elements heavier than iron relative to the number of Ni atoms present in the ejecta. This step ensures that the abundance ratios are consistent with known solar data while accounting for the specific conditions in the jet.

Research on mass ejection and nucleosynthesis in BNS mergers indicates that the temperature of the dynamical ejecta typically exceeds $10 \text{GK}$ \cite{arXiv:2205.05557}, equivalent to approximately 0.86 MeV. 
Moreover, during the merger process, shock heating can generate temperatures as high as $100 \text{MeV}$ at the contact layer between the colliding stars.
The collision dislodges neutron-rich material, which is subsequently carried into the jet by the neutrino cooling winds and magnetic fields from the polar region.
The post-merger ejecta undergo heating due to temperature inversion caused by differential rotation, reaching temperatures around $40 \text{MeV}$. 
Isomeric states of the r-processes elements created in the ejecta of neutron star mergers could be populated due to these high temperatures, according to the Boltzmann distribution:
\begin{equation}
N_{\text{i,0}}= N_{\text{count}} e^{-\frac{\Delta E}{kT}}.
\label{eq:boltzmann}
\end{equation}
Here, $\Delta E$ represents the energy difference between the isomeric and ground states, $k$ is the Boltzmann constant, and $T$ denotes the temperature of the ejecta.
Besides nuclear excitation at high temperatures, the isomeric states can also be activated through absorption of $\gamma$-ray photons, or interaction with ultraviolet and X-rays, for nuclei moving at relativistic speeds. 
In a future work, after a more detailed analysis of the properties of the matter in the jet, we will refine the number of isomers at specific excitation energies.

\begin{itemize}[nosep]
\item 
In the cases where an isotope admits an isomer that has multiple decay pathways, the energy contribution from each decay is weighted according to the probability $P_i$ of that specific pathway, 
\begin{equation}
P_i = \frac{1}{N} \frac{E_{\gamma, i}}{\langle{E_{\gamma}}\rangle},
\label{eq:probability}
\end{equation}
where $N$ is the total number of isomeric decays allowed for a particular istope in isomeric state, $i$ is the decay path, $E_{\gamma, i}$ the energy of the decay path, and $\langle{E_{\gamma}}\rangle$ the mean value of the energies.
This holds, because the probability of each decay path is proportional to its $\gamma$-ray energy \cite{arXiv:2208.01028}.
\item 
We calculate the number of $\gamma$-ray radiation events $\Delta N _i$ occurring within a chosen time interval $\Delta t$ for each isomer identified, using the decay law:
\begin{equation}
\Delta N_i = N_{i,0} e^{-\lambda t} \left(1 - e^{-\lambda \Delta t}\right)
\end{equation}
Here, $\lambda = \ln(2)/T_{1/2}$ is the decay constant, with $T_{1/2}$ being the half-life of the isomer.
We focus on isomers with terrestrial half-life between $t_0-t_r \le T_{1/2}  \le t_j$, assuming that the $\gamma$-rays emitted by isomeric de-excitation between the end of the r-process and the start of the jet will be reabsorbed.
%\item
%For each isomer, we multiply the number of decays $\Delta N_i$ by the $\gamma$-ray energy released per decay. This gives us the total energy emitted in the interval $\Delta t$. This calculation is repeated for each isomer.
\item
Finally, we combine the number of decays (or counts) $\Delta N_i$ for all the isomers within a given time interval to obtain the cumulative output from all isomeric decays for each specific de-excitation energy.
\end{itemize}

To construct the light curve  that captures the contribution of isomeric $\gamma$-ray de-excitation to the temporal evolution of the GRB's intensity, we proceed as follows:
\begin{itemize}[nosep]
\item  For each isomer, we multiply the number of counts $\Delta N_i$ by the $\gamma$-ray energy released per decay for consecutive time intervals, to obtain the evolution of the intensity in time, in increments of $\Delta t$.
%This gives us the total energy emitted in the interval $\Delta t$.
%Once we calculate the number of isomer de-excitation (or counts) for consecutive time intervals, in increments of $\Delta t$. and divide by $\Delta t$ 
\item Then, we calculate the luminosity emitted for each time bin, by summing the emitted intensities for all the isomers and dividing by $\Delta t$. %, in essence reducing the spectrum on each time interval to a point.
%, providing a simplified yet insightful view of how the energy changes in time. %light curve. 
\item We calculate the specified emission area as $A = \Omega_j r^2$, where $r$ is the estimated radius at which the jet is emitted, and its collimation angle is $\theta_j = (1/r)^p$, where $p\approx 0.22$ \cite{arXiv:1912.00057}. 
\item Lastly, we divide the luminosity calculated by the relevant area, to obtain the flux of $\gamma$-ray. 
%This method is particularly useful in GRB studies, as the peak energy often varies significantly over time, offering key insights into the dynamics of the event.
\end{itemize}

We have outlined the methods employed for building an interactive tool for analyzing data related to the complex process behind the contribution of isomeric transitions to the GRB spectrum, and to construct the light curve that captures the contribution of isomers to the temporal evolution of the GRB's intensity.
%process used to build an interactive web page and the 

\section{Results}

To facilitate the data analysis of nuclear isomers with  $\gamma$-ray emission relevant to GRBs, we developed an interactive web interface using \href{https://streamlit.io/}{Streamlit}, an open-source Python library designed for creating shareable data applications. Our application, retrievable at \href{https://isomersearchengine.streamlit.app}{https://isomersearchengine.streamlit.app},  allows users to filter the data according to specific criteria through interactive sidebars. It also provides functionalities to plot relevant graphs and download the selected data for further analysis. This streamlined approach significantly enhances the efficiency and accessibility of our nuclear isomer data analysis tool.

We started our data analysis by estimating the amount of matter containing r-process elements that can contribute to the overall $\gamma$-ray decays occurring within the jets from neutron star mergers, a crucial quantity necessary to calculate the isotope count. 
We considered a jet of typical geometry, with an initial half opening angle of approximately $\theta_0 = 20^\circ$ when the jet forms \cite{arXiv:2312.01074}.
Using eq.(\ref{eq:Meff}) with this value for $\theta_0$, we found that the effective mass in the jet is $M_{\text{eff}}\approx 6\%M_{\text{ejecta}}$.
In this case, the early ejecta contributes minimally, with a mass of maximum $6 \times 10^{-4} M_{\odot}$.
The more significant contribution, of $6 \times 10^{-3} M_{\odot}$ arises from the post-merger ejecta.
Therefore, adding the mass of the magnetized wind to the dynamic and post-merger ejecta expected to impact the jet, we obtain an effective mass $M_{\text{eff}} \approx 10^{-2} M_{\odot}$.
This effective mass contains $1.2 \times 10^{55}\text{baryons}$, among which we assume that only $10\%$ contribute to the formation of heavy elements \cite{arXiv:1710.05836}.
This approach allowed us to determine the amount of matter containing r-process elements that can contribute to the overall decay processes occurring within the jet.
Our assumption is based on the premise that the ejecta is spherically-symmetric and that relevant $\gamma$-ray emission influencing the GRB's energetic output comes from the matter within this conical section of the jet.
This model provides only an approximate estimate of the matter distribution within the jet.  
A detailed investigation of the matter distribution as a function of the angle, which is essential for a more precise understanding, is beyond the scope of this study and will be the subject of future work.

We determine the initial number of isotopes in $M_{\text{eff}}$ and select among these heavy elements the isotopes that admit isomers with $100\%$ isomeric transitions, then calculate their abundance.
Our analysis of the ejecta impacting the jet revealed that approximately $35\%$ of the mass comes from the wind, about $5\%$ originates from tidal interactions, and the remaining $60\%$ is attributed to the outflow. 
Considering the different contributions to the ejecta and their respective temperatures, the average temperature of the ejecta is approximately $T = 59 \text{MeV}$.
We infer from eq.\ref{eq:boltzmann} that for this average temperature, more than $96\%$ of the heavy isotopes formed in the ejecta will have their isomeric energy levels populated. 
This provides only a rough estimate, because we did not account for the isomers activated through other processes.
We mediate the number of isomers for each emitted $\gamma$-ray energy with the probability of de-excitation using eq.(\ref{eq:probability}).

We plot in Fig.\ref{fig:Figure1} the abundances of heavy elements, starting with strontium, as a function of the atomic number $Z$ according to the predicted solar system abundances, first for the isotopes (Fig.\ref{fig:Figure1a}), and then for their corresponding isomers (Fig.\ref{fig:Figure1b}). 
These are shown relative to the abundance of $_{234}^{92}U$, which has the lowest number of atoms ($_{234}^{92}U = 2.51662\times 10^{44}$ atoms).
The number of different isotopes for each $Z$ is indicated within each bar, and the legend on the right lists the type of atoms. 
Strontium is the most abundant isotope, followed by xenon and lead, while in isomeric state only xenon and lead keep their contribution.

In Fig. \ref{fig:Figure2a}, we present the abundance of isomers on a logarithmic scale, plotted against the emitted $\gamma$-ray energy and the logarithm of the de-excitation time, providing an intuitive overview of all the isomers created in a BNS merger ejecta. 
This is complemented by the complete de-excitation spectra shown in Fig. \ref{fig:Figure2b}. 
We observe that the majority of isomers de-excite in less than ($10 \mu\text{s}$) and have their de-excitation energies below $500\text{keV}$. These emissions are likely reabsorbed by the medium within the jet, and are most likely not contributing to its overall energy budget.
Observational data and models of GRBs indicate that the peak $\gamma$-ray energy ranges between $500\text{keV}$ and $2\text{MeV}$. Consequently, isomers emitting within this energy range merit further investigation.
In Fig. \ref{fig:Figure3a}, we present the isomers energy spectrum as cumulative counts of de-excitations occurring within a time interval of $3\text{s}$, highlighting all the isomers with the largest de-excitations counts, over $10^{50}$. 
We note that $_{132}Xe$, with a de-excitation energy of $538.2$ provides the most substantial contribution to the overall emission.
We further narrow down our selection to isomers with de-excitation times exceeding $10 \mu\text{s}$ and energies greater than $500\text{keV}$, and present them in Fig. \ref{fig:Figure3b}, highlighting now the five isomers with the largest de-excitations counts.
We list the identified isomers in Table \ref{tab2}, detailing their lifetimes, atomic mass number, and their emitted $\gamma$-ray energies. 

We proceed to illustrate in Fig. \ref{fig:Figure4a} the ${\gamma}$-ray intensity emitted over time, by the de-excitation of the five isomers selected, from $0$ to $3\text{s}$, calculated as $Intensity(t) = E_{\gamma} \times  {Counts}(t)$.
We observe minimal intensity contribution from the $_{206}\text{Pb}$ isomer, while $_{132}\text{Xe}$ peaks around $t \approx 5 \text{ms}$ before steeply declining. Both of these isomers have de-excitation energies around $500 \text{keV}$. 
Factoring in the electron-positron annihilation process, expected to be frequent within the jet and releasing $1.022\text{MeV}$, it becomes challenging to differentiate the signal from these isomers from that of the annihilation process.
Consequently, our focus will shift to isomers with de-excitation energies above $1\text{MeV}$ and de-excitation times of seconds, specifically $_{89}\text{Y}$, $_{207}\text{Pb}$, and $_{90}\text{Zr}$, as these characteristics make their signals more distinguishable and relevant to our study.
In Fig. \ref{fig:Figure4b}, we show the intensity evolution of these three isomers. 
Notably, all three exhibit a peak at approximately $0.15 \text{s}$, followed by a rapid decline. Based on these observations, we identify $_{90}\text{Zr}$, $_{207}\text{Pb}$ and $_{89}\text{Y}$ as prime candidates to account for the hard prompt $\gamma$-ray emission observed in GRBs.

Lastly, we construct the light curves, which requires estimating the radius at which the jet originates from the central engine, which is still uncertain \cite{arXiv:2306.16795}. 
In Fig. \ref{fig:Figure5}, we compare the evolution of the total luminosity over time from all isomers against the luminosity specifically from the three selected isomers, displayed on a logarithmic scale. Initially, there is a notable burst in luminosity, attributed to the rapid de-excitation of isomers with lifetimes under $1 \mu \text{s}$. 
However, after about $0.1\text{s}$, the emission comes prominently from $_{132}\text{Xe}$ and the three key isomers, $_{90}\text{Zr}$, $_{207}\text{Pb}$, and $_{89}\text{Y}$, indicating their potential significant contribution to the jet. We note that the peak luminosity emitted by the selected isomers is $L^{\text{max}}_{\text{(Pb, Zr,Y)}}\approx 7.15\times 10^{44} \text{erg/s}$ lower than that of the peak luminosity of typical GRB ($L^{\text{max}}_{\text{GRB}}\approx 10^{50}\text{erg/s}$). 
This value represents the `true' luminosity of the $\gamma$-ray photons that are emitted by isomeric transitions  within the jet, and no further adjustment with the beaming factor is necessary. 
To calculate the flux at the GRB emission (in CGS units), we adopt the assumption that the jet forms at a distance of $5 \times 10^6\text{cm}$ from the central engine \cite{arXiv:1907.07599}, with a half-opening angle of $\theta_0 = 20^\circ$ and the emission occurs once the jet extends to about $10^{9}\text{cm}$ from the central engine. By this stage, the collimation angle of the jet is approximately $6.23^\circ$, covering a surface area of $A_{\text{jet}}=3.72 \times 10^{16}\text{cm}^2$. For these values, we obtain a peak flux of $F^{\text{max}}_{\text{(Pb, Zr,Y)}}\approx 1.92\times 10^{28} \text{erg/(s}\cdot \text{cm}^2)$.

Our findings present an analysis of isomeric abundances and their energy spectrum within the jet.
We identify the top three relevant isomers with the potential to influence GRB gamma-ray emission. 
By examining the luminosity and flux generated by these isomers, we provide a foundational understanding of their contribution to the prompt emission phase of GRBs. 
It is important to note that our calculations while not accounting for the Doppler boost, likely present an upper-bound estimate. This is because we have assumed that all elements within the jet are in isomeric states that de-excite exclusively through isomeric transitions. Additionally, we treated the ejecta as isotropic and structurally uniform, without considering the diverse components and their respective r-process element abundances. 
This study paves the way for deeper investigations into the complex dynamics of these cosmic phenomena.

\section{Conclusion}

In this study, we relied on the knowledge that neutron star mergers play a crucial role in creating elements heavier than iron through r-process nucleosynthesis. 
The starting point of our investigation was the binary neutron star (BNS) merger GW170817, a milestone event observed both in gravitational waves and electromagnetic radiation. The prompt gamma-ray emission spectrum of the accompanying gamma ray burst (GRB 170817A) continues to be an open question.
We proposed a novel idea, namely that the $\gamma$-ray spectrum of such GRBs may include contributions from $\gamma$-ray de-excitations due to isomeric transitions.

Our research starts with a comprehensive examination of the current understanding of GRB structure, coupled with an investigation into r-process nucleosynthesis during neutron star collisions. We make a case for the addition of isomers within these astrophysical phenomena.
To investigate the role played by isomeric transitions within the GRB emission, we created an interactive web page designed to facilitate a thorough analysis of their potential impact on the GRB $\gamma$-ray spectra. This platform allows for interactive data filtering, detailed visualization of radiation spectra, and light curve modeling. 
We began by selecting representative isomers and estimating their initial quantities, using known solar element abundances and factoring in the quantity of matter expected to influence on gamma-ray production.
Subsequently, we computed the number of gamma-ray radiation events for each isomer. This data was then used to construct both the radiation spectrum and the light curve, tailored specifically to the time interval of GRB 170817A. 
We identified three isomers, $_{90}\text{Zr}$, $_{207}\text{Pb}$, and $_{89}\text{Y}$, whose abundance, de-excitation energy, and lifetime make them prime candidates for contributing to the prompt GRB spectrum.
This approach provides a comprehensive method for examining the $\gamma$-ray characteristics of GRBs from similar astrophysical events.

Moving forward, our next goal is to refine our methods and compare the theoretical spectra and light curves predicted by our model against actual observations of GRBs from r-process sites. This comparison will be crucial in testing our assumptions and validating our model, thus deepening our understanding of the GRB emission spectra. 
In upcoming work, we plan to expand our selection of nuclear isomers to include elements lighter than strontium, and to pursue a more detailed analysis of the matter distribution and temperature in the jet. These improvements will allow us to achieve a more precise calculation of isomeric abundance in the major production sites of elements, apply our model to long GRBs, and incorporate these findings into astrophysical simulations. This last step will provide us with accurate calculations of isomeric abundance in astrophysical r-processes and will enable us to identify the precise contribution of isomeric transitions to the $\gamma$-ray signatures of GRBs, thus enhancing our understanding these astrophysical events. 

%In conclusion, our research not only advances the hypothesis of isomeric contributions to GRB gamma-ray spectra but also introduces a new method for investigating these contributions. 
%Future detections of GW/GRB events will provide further insights into the GRB spectra from neutron star mergers and the correlation between the peak of the GW signal and the emission of $\gamma$-rays. Such data will offer further opportunities to test the predictions of our model and shed light on the underlying mechanisms of GRBs, enhancing our understanding of high-energy astrophysical events.
%In a future work, we will pursue a more detailed analysis of matter distribution and temperature in the jet, and we will employ the Boltzmann distribution to accurately calculate the exact number of isomers at specific excitation energies. %This approach will enable us to achieve a more precise understanding of the isomeric states within these astrophysical phenomena.
%This assumption is key in understanding the initial conditions and nuclear processes occurring within these extreme and dynamic environments

%By comparing our model with real-world data, we aim to refine the precision of our predictions and contribute to obtain a wider knowledge  of these fascinating cosmic events.
%broadening the understanding 

%For additional requirements for specific article types and further information please refer to the individual Frontiers journal pages

\section*{Conflict of Interest Statement}
%All financial, commercial or other relationships that might be perceived by the academic community as representing a potential conflict of interest must be disclosed. If no such relationship exists, authors will be asked to confirm the following statement: 
The authors declare that the research was conducted in the absence of any commercial or financial relationships that could be construed as a potential conflict of interest.

\section*{Author Contributions}
%The Author Contributions section is mandatory for all articles, including articles by sole authors. If an appropriate statement is not provided on submission, a standard one will be inserted during the production process. The Author Contributions statement must describe the contributions of individual authors referred to by their initials and, in doing so, all authors agree to be accountable for the content of the work. Please see  \href{https://www.frontiersin.org/guidelines/policies-and-publication-ethics#authorship-and-author-responsibilities}{here} for full authorship criteria.
%authors contributed to the writing and editing of the manuscript. JHG initiated and led the project,  led the writing and editing of the manuscript, and produced Figures ...
%CLF led the data reduction 
MCBH initiated and led the project, led the writing and editing of the manuscript, and produced the figures. 
JIP led the data reduction, contributed to analysis and built the web page.
Both authors listed significantly contributed directly and intellectually to this work and give their approval for its publication.
%MCBH has written this article. The references provided offer a comprehensive coverage, yet they do not list all available sources.
%JIP has produced all figures himself from publicly available data and from the data analysis. 

\section*{Funding}
%Details of all funding sources should be provided, including grant numbers if applicable. Please ensure to add all necessary funding information, as after publication this is no longer possible.
This research was made possible in part by the NASA Established Program to Stimulate Competitive Research, Grant No. 80NSSC22M0027 and by NSF grant No. PHY-1748958 to Kavli Institute for Theoretical Physics.
JIP also acknowledges support from the Summer Undergraduate Research Experience (SURE) Program funded by the West Virginia Research Challenge Fund.

%\section*{Acknowledgments}
%This is a short text to acknowledge the contributions of specific colleagues, institutions, or agencies that aided the efforts of the authors.

%\section*{Supplemental Data}
% \href{https://www.frontiersin.org/guidelines/author-guidelines#supplementary-material}{Supplementary Material} should be uploaded separately on submission, if there are Supplementary Figures, please include the caption in the same file as the figure. LaTeX Supplementary Material templates can be found in the Frontiers LaTeX folder.

\section*{Data Availability Statement}
The interactive website application can be accessed at the address \href{https://isomersearchengine.streamlit.app}{isomersearchengine.streamlit.app}.
The python code developed and the datasets generated for this study are available at the gitlab repository \href{https://github.com/Powell222/Isomer_Search_Engine}{$\text{github.com/Powell222/Isomer\_Search\_Engine}$}.
%Access to the development code will be granted upon a reasonable request directed to the corresponding authors.
% Please see the availability of data guidelines for more information, at https://www.frontiersin.org/guidelines/policies-and-publication-ethics#materials-and-data-policies

%\bibliographystyle{Frontiers-Harvard} 
%  Many Frontiers journals use the Harvard referencing system (Author-date), to find the style and resources for the journal you are submitting to: https://zendesk.frontiersin.org/hc/en-us/articles/360017860337-Frontiers-Reference-Styles-by-Journal. For Humanities and Social Sciences articles please include page numbers in the in-text citations 
\bibliographystyle{Frontiers-Vancouver} 
% Many Frontiers journals use the numbered referencing system, to find the style and resources for the journal you are submitting to: https://zendesk.frontiersin.org/hc/en-us/articles/360017860337-Frontiers-Reference-Styles-by-Journal
%\bibliographystyle{frontiersinHLTH&FPHY} % for Health, Physics and Mathematics articles
\bibliography{references}

%%% Make sure to upload the bib file along with the tex file and PDF
%%% Please see the test.bib file for some examples of references

\section*{Figures and Tables}

\begin{table}[h]
\caption{The estimated fractions of each element created in merging neutron stars \cite{johnson}}\label{tab1}%
\begin{tabular}{@{}llll@{}}
\toprule
Fraction & Element \\
\midrule
$0.25$ & Se, Br, Kr, Rb, Sr, Y, Zr, Nb, Sn, Ba, Ce, Tl, Bi  \\
$0.50$ & Mo, Pd, Cd, Te, La, Pr, Nd, Hf, Ta, W, Hg  \\
$0.75$ &  Ru, In, Sb, Sm, Yb, Lu \\
$1.00$ & Rh, Ag, I, Xe, Cs,  Eu, Gd, Tb, Dy, Ho, Er, Tm, Re, Os, Ir, Pt, Au, Pb, Th, U  \\
\botrule
\end{tabular}
\end{table}

\begin{table}[h]
\caption{The isomers within the selected criteria.}\label{tab2}%
\begin{tabular}{@{}llll@{}}
\toprule
Fraction & Element & Lifetime (s) & Energy (keV)  \\
\midrule
$206$ & Pb & $1.25\times 10^{-4}$ & $516.18$  \\
$132$ & Xe & $8.37 \times 10^{-3}$ & $538.2$  \\
$89$ & Y & $15.66 \times 10^{-3}$ & $908.96$  \\
$207$ & Pb & $0.805\times 10^{-4}$ & $1063.656$  \\
$90$ & Zr & $0.809$ & $2319$  \\
\botrule
\end{tabular}
\end{table}

%%%%%%%%%% Figure1 %%%%%%%%%%%
\setcounter{figure}{1}
\setcounter{subfigure}{0}
\setlength{\abovecaptionskip}{-5pt}
\begin{subfigure}
\setcounter{figure}{1}
\setcounter{subfigure}{0}
    \centering
    \begin{minipage}[b]{1.0\textwidth}
        \includegraphics[width=\linewidth]{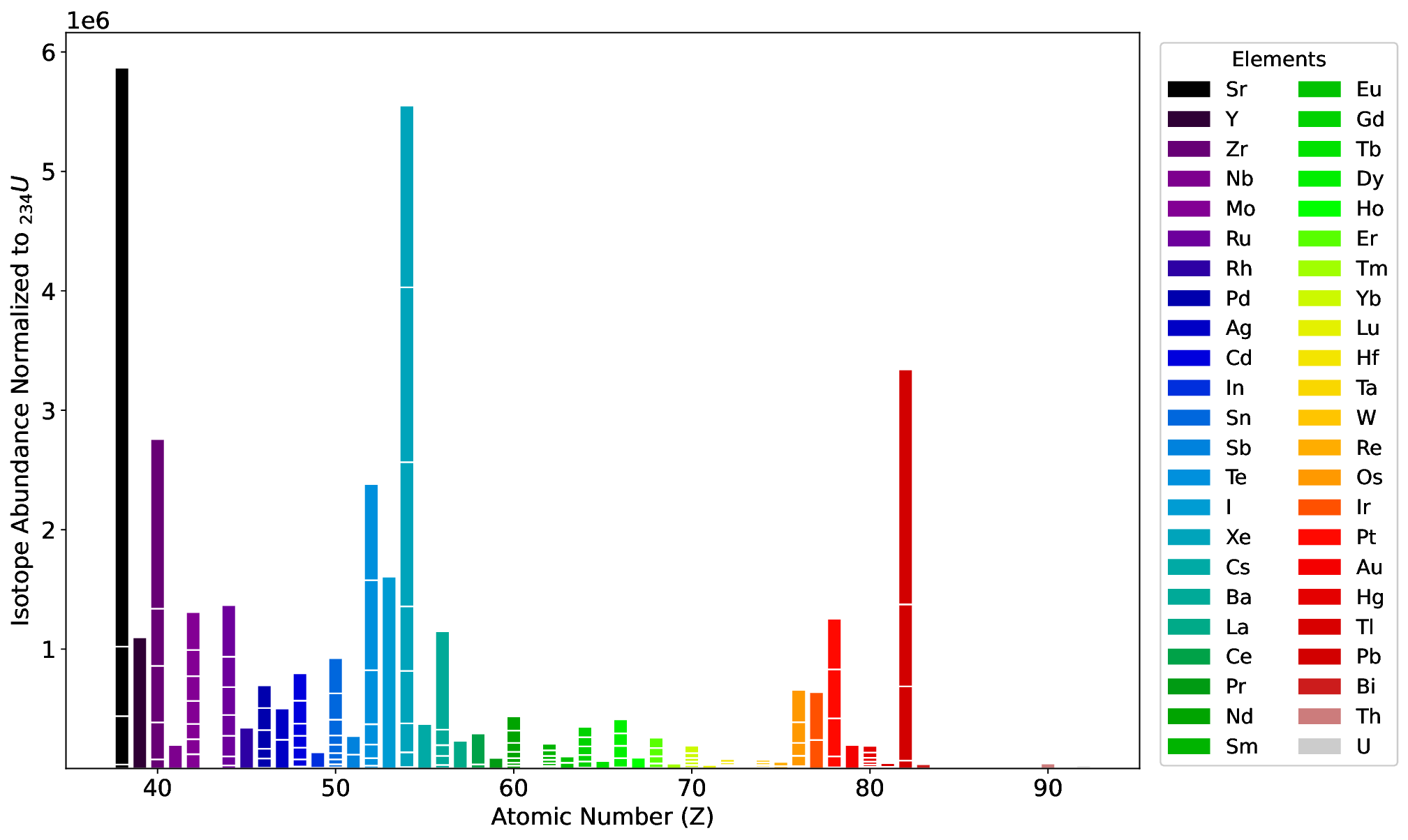}
        \caption{Isotopes abundance per atom of $_{234}^{92}U$ created in BNS merger ejecta. Note: vertical axis $\times 10^6$.}
        \label{fig:Figure1a}
    \end{minipage}  
    
%\vspace{+10pt}   
\setcounter{figure}{1}
\setcounter{subfigure}{1}
\setlength{\abovecaptionskip}{-5pt}
    \begin{minipage}[b]{1.0\textwidth}
    \vspace{+15pt}
        \includegraphics[width=\linewidth]{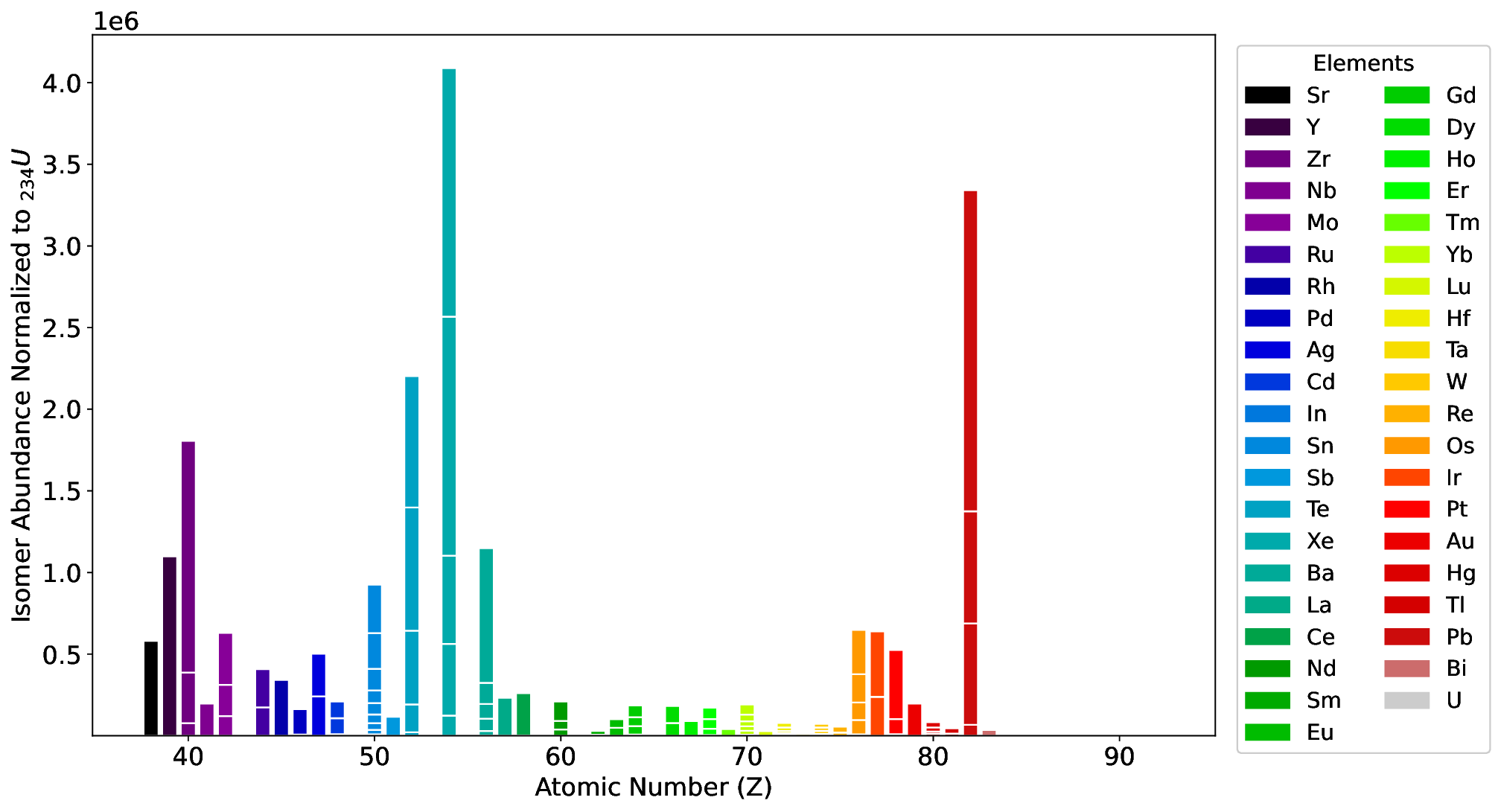}
        \caption{Isomer abundance per atom of $_{234}^{92}U$ created BNS merger ejecta. Note: vertical axis $\times 10^6$.}
        \label{fig:Figure1b}
    \end{minipage}

\setcounter{figure}{1}
\setcounter{subfigure}{-1}
\setlength{\abovecaptionskip}{+10pt}
    \caption{Abundances of heavy elements in BNS merger ejecta starting with strontium, plotted against atomic number $Z$, normalized to the lowest number of atoms ($_{234}^{92}U$), and stacked by atomic mass number.}
    \label{fig:Figure1}
\end{subfigure}

%%%%%%%%%%%Figure2 %%%%%%%%%%%%
\setcounter{figure}{2}
\setcounter{subfigure}{0}
\setlength{\abovecaptionskip}{-15pt}
\begin{subfigure}
\setcounter{figure}{2}
\setcounter{subfigure}{0}
    \centering
    \begin{minipage}[b]{1.0\textwidth}
        \includegraphics[width=\linewidth]{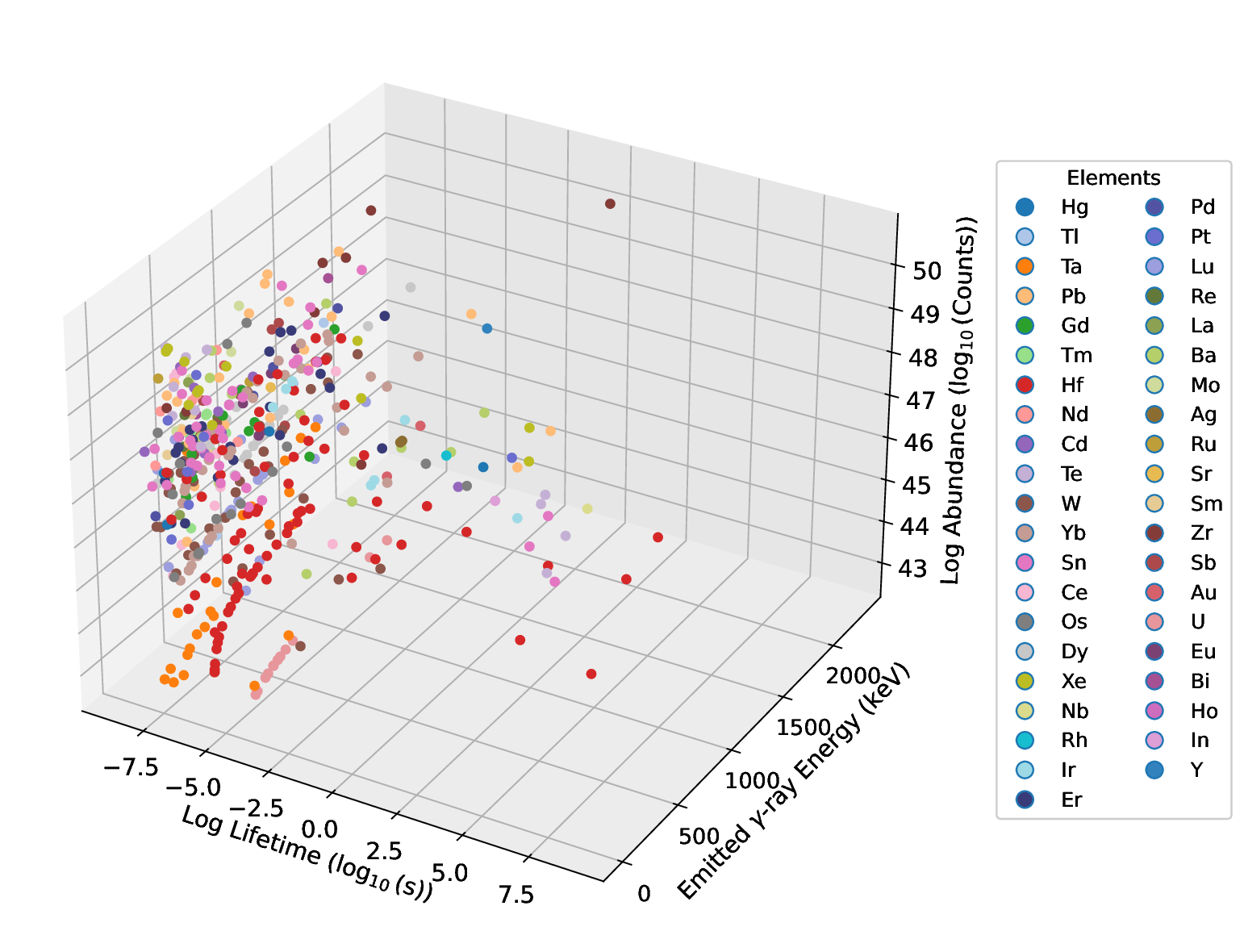}
        \caption{Log isomer abundance vs log lifetime and $\gamma$-ray emitted energy.}
        \label{fig:Figure2a}
    \end{minipage}  
    
%\vspace{+10pt}   
\setcounter{figure}{2}
\setcounter{subfigure}{1}
\setlength{\abovecaptionskip}{-5pt}
    \begin{minipage}[b]{1.0\textwidth}
    \vspace{+10pt}
        \includegraphics[width=\linewidth]{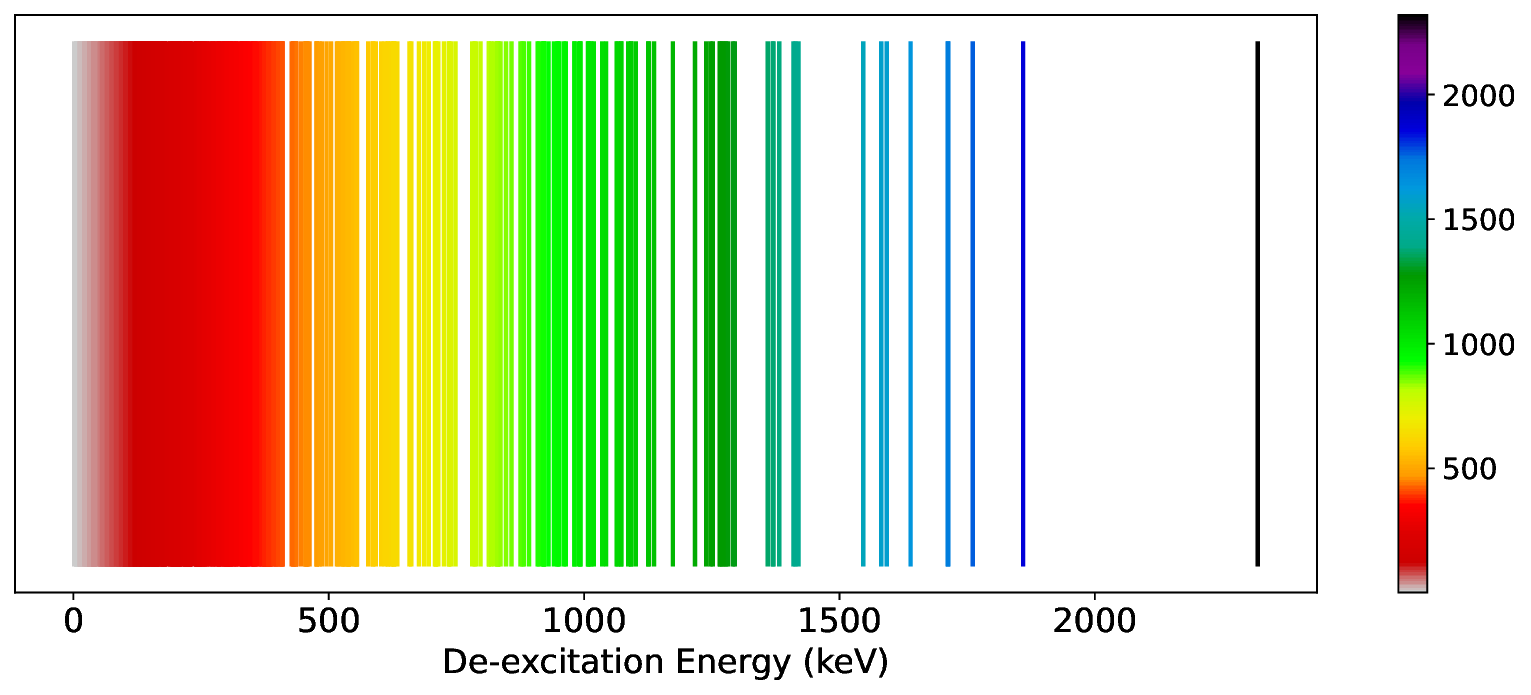}
        \caption{The emission spectrum of all the nuclear isomers in the BNS ejecta}
        \label{fig:Figure2b}
    \end{minipage}

\setcounter{figure}{2}
\setcounter{subfigure}{-1}
\setlength{\abovecaptionskip}{+10pt}
    \caption{\textbf{(A)} Logarithmic abundance of isomers in the jet vs logarithm of the lifetime ($T_{1/2}$) and the corresponding emitted $E_{\gamma}$. \textbf{(B)} The spectra of the emitted $\gamma$-ray radiation for all the isomers in the jet.}
    \label{fig: Figure2}
\end{subfigure}

%%%%%%%%% Figure 3 %%%%%%%%%%%%%%
\setcounter{figure}{3}
\setcounter{subfigure}{0}
\setlength{\abovecaptionskip}{-5pt}
\begin{subfigure}
\setcounter{figure}{3}
\setcounter{subfigure}{0}
    \centering
    \begin{minipage}[b]{\textwidth}
        \includegraphics[width=0.9\linewidth]{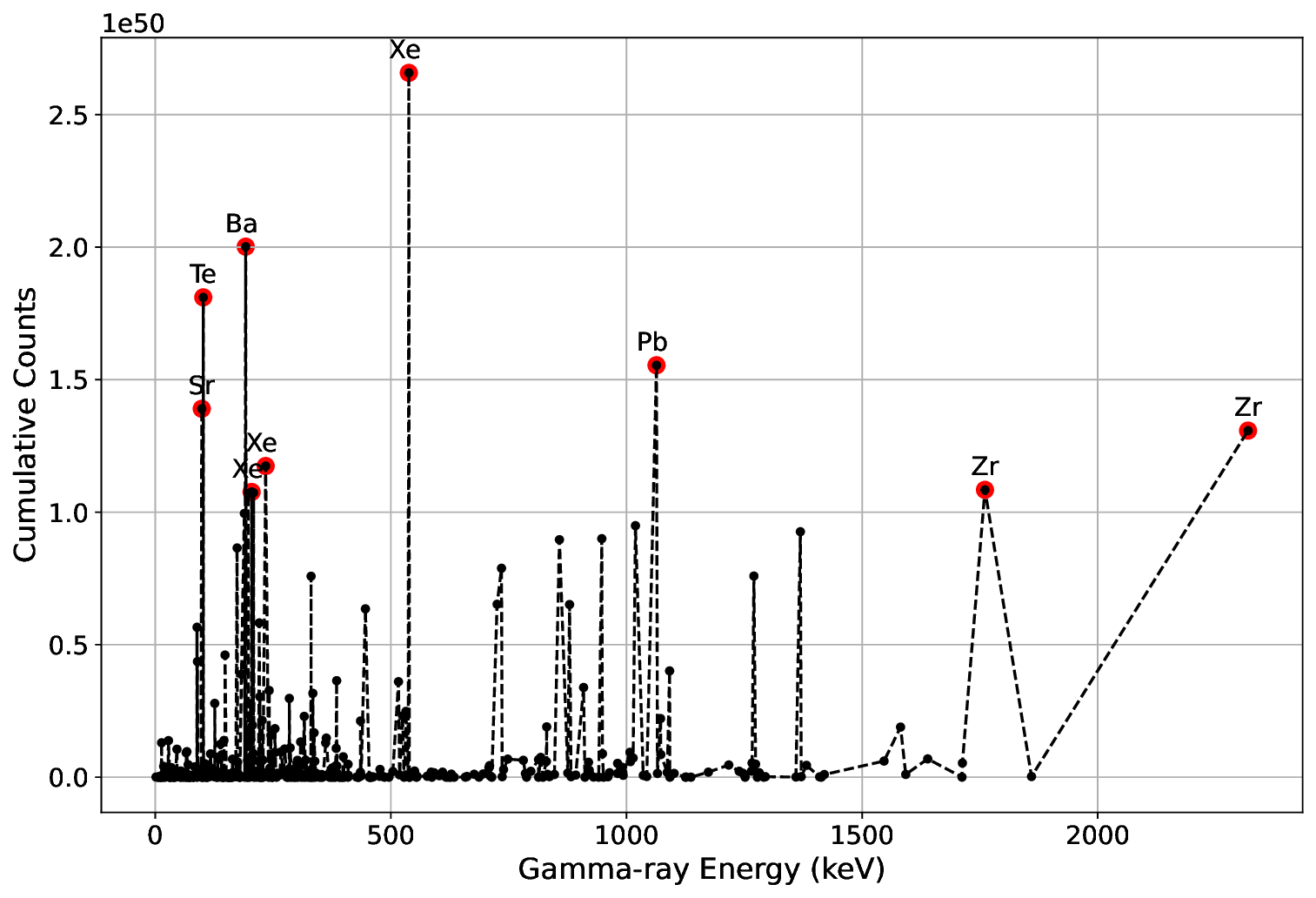}
        \caption{Isomer de-excitations spectra for $t\le 3\text{s}$, with top elements highlighted. Vertical axis $\times 10^{50}$}
        \label{fig:Figure3a}
    \end{minipage}  
    
%\vspace{+10pt}   
\setcounter{figure}{3}
\setcounter{subfigure}{1}
\setlength{\abovecaptionskip}{-5pt}
    \begin{minipage}[b]{0.9\textwidth}
    \vspace{+10pt}
        \includegraphics[width=\linewidth]{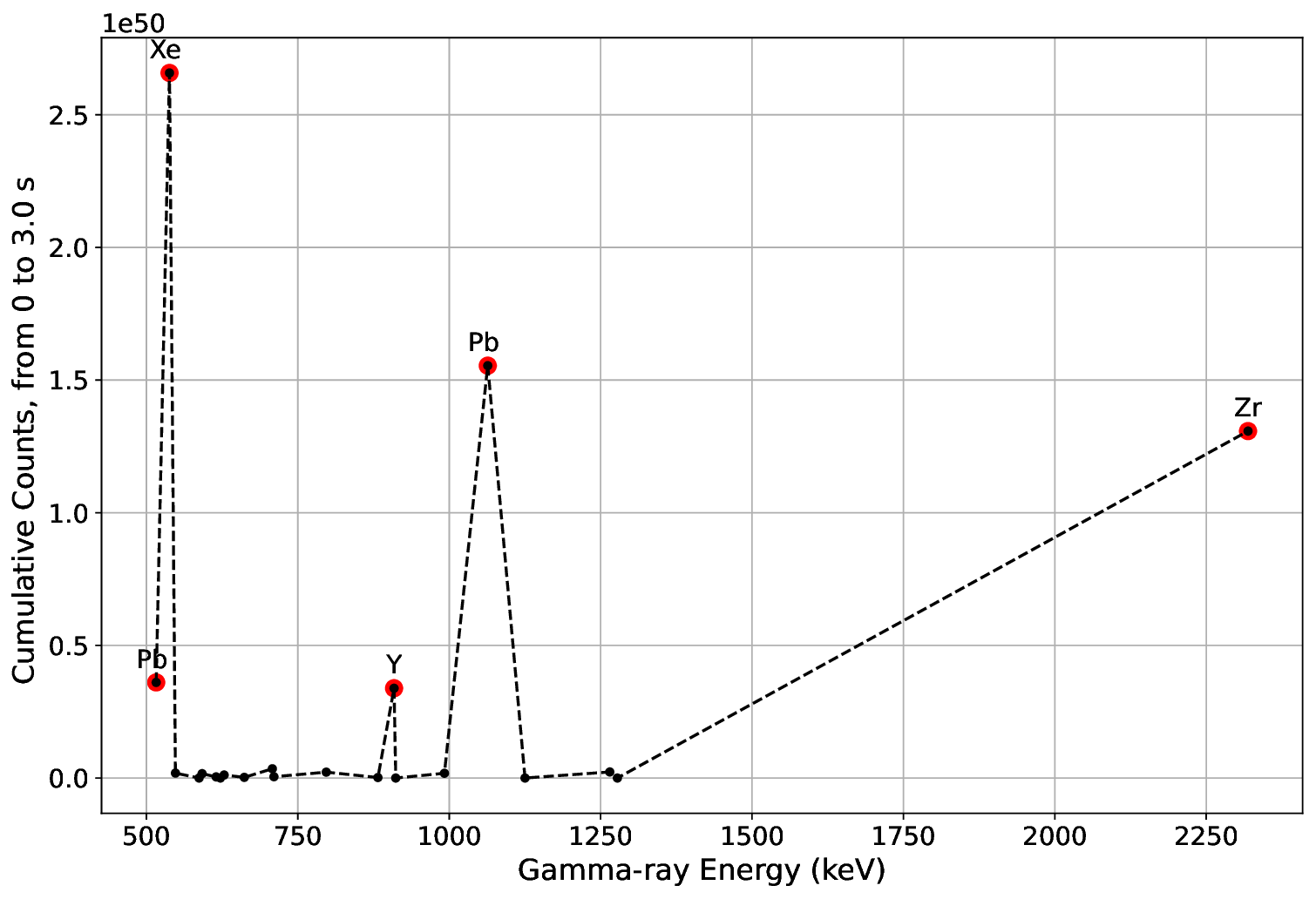}
        \caption{Selected isomer $\gamma$-ray spectra ($T_{1/2} \ge 10 \mu\text{s}$, $E_{\gamma}\ge 500\text{keV}$). Vertical axis $\times 10^{50}$.}
        \label{fig:Figure3b}
    \end{minipage}

\setcounter{figure}{3}
\setcounter{subfigure}{-1}
\setlength{\abovecaptionskip}{+10pt}
    \caption{Isomer de-excitations spectra, represented as cumulative counts vs. $\gamma$-ray energy, $t\le 3\text{s}$.}
    \label{fig:Figure3}
\end{subfigure}

%%%%%%%%% Figure 4 %%%%%%%%%%%%%%
\setcounter{figure}{4}
\setcounter{subfigure}{0}
\setlength{\abovecaptionskip}{-5pt}
\begin{subfigure}
\setcounter{figure}{4}
\setcounter{subfigure}{0}
    \centering
    \begin{minipage}[b]{0.9\textwidth}
        \includegraphics[width=\linewidth]{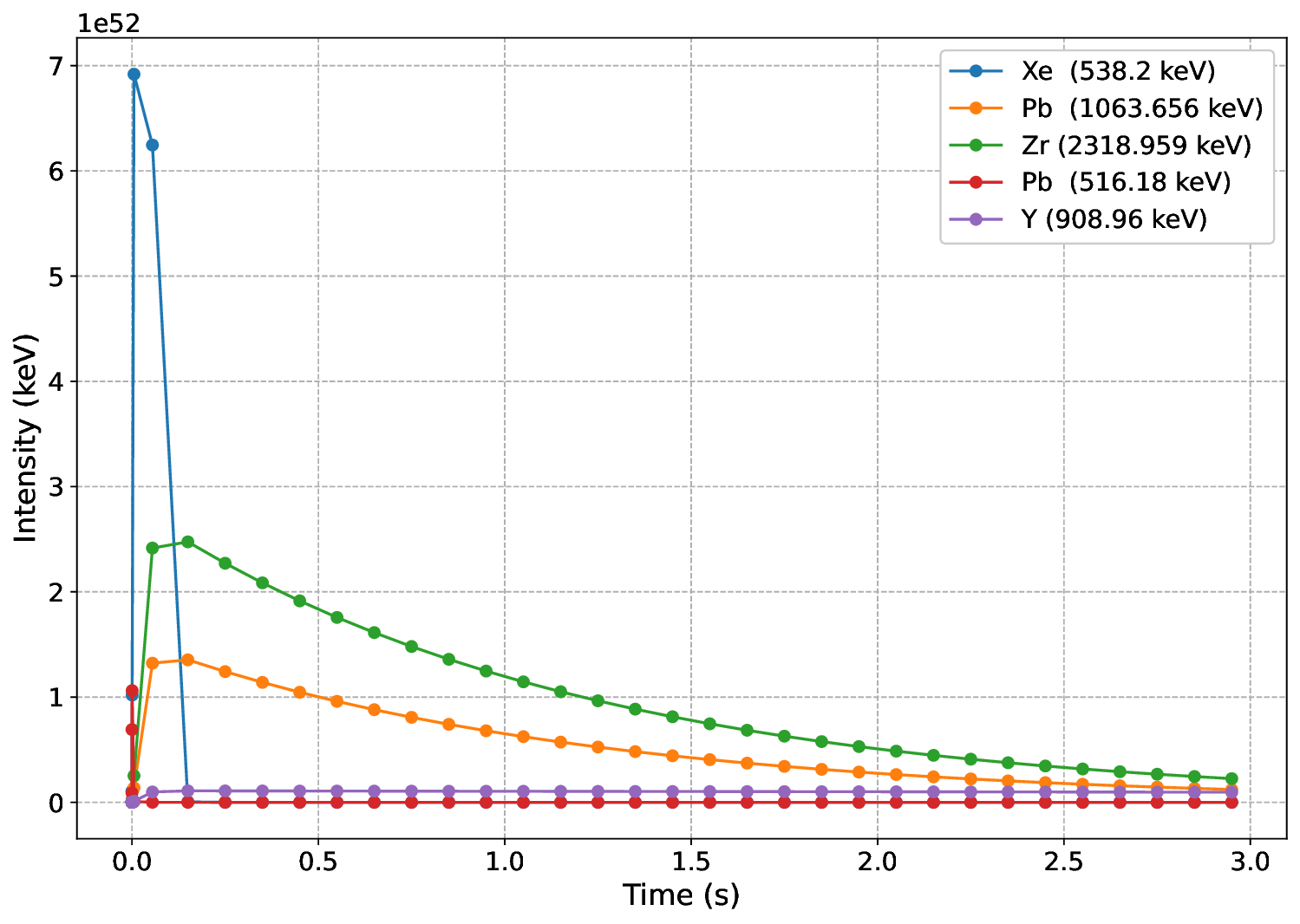}
        \caption{Evolution of the intensity for top 5 isomers, with $E_{\gamma}\ge 500\text{keV}$. Vertical axis $\times 10^{52}$.}
        \label{fig:Figure4a}
    \end{minipage}  
    
%\vspace{+10pt}   
\setcounter{figure}{4}
\setcounter{subfigure}{1}
\setlength{\abovecaptionskip}{-5pt}
    \begin{minipage}[b]{0.9\textwidth}
    \vspace{+15pt}
        \includegraphics[width=\linewidth]{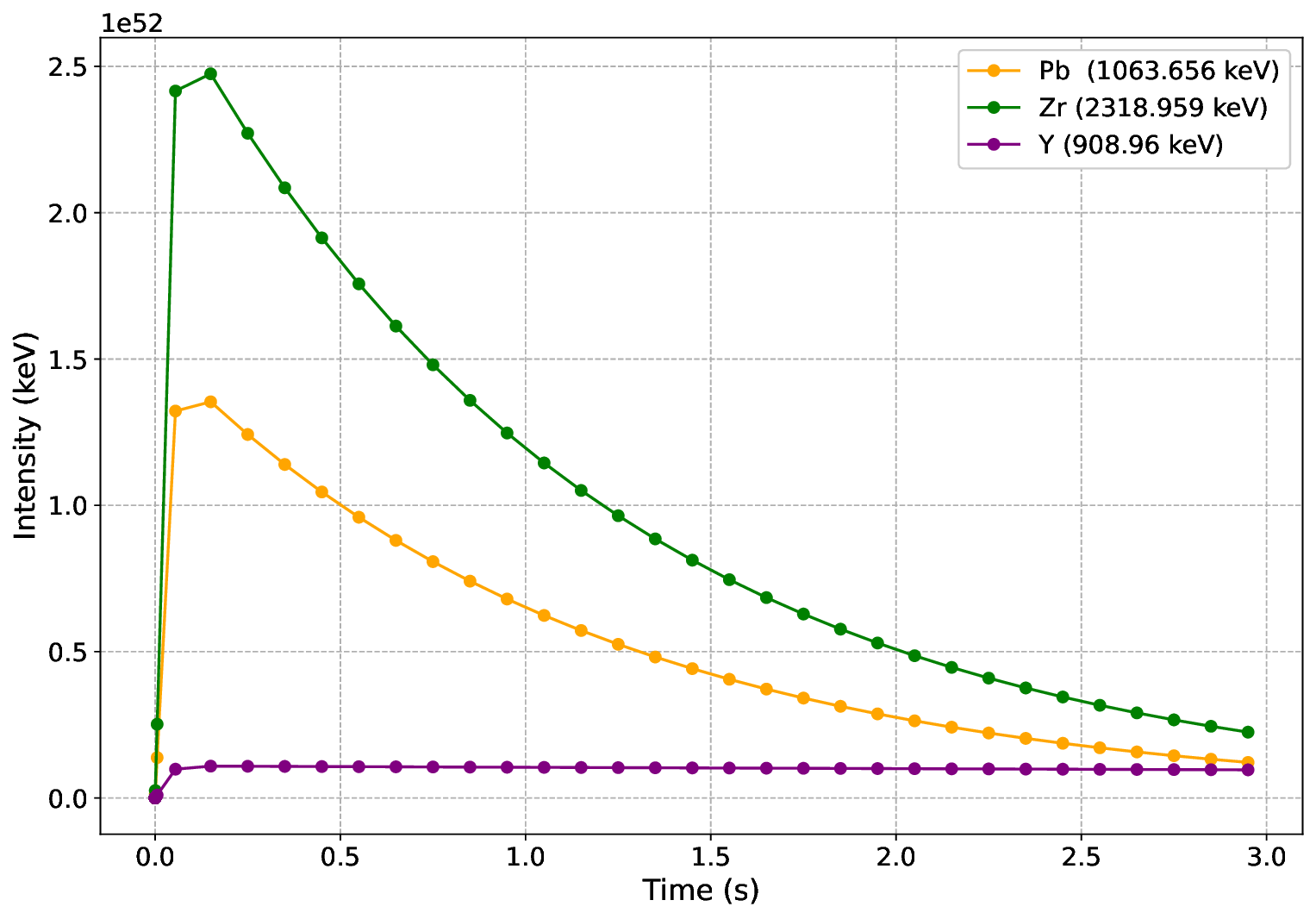}
        \caption{Evolution of the intensity for top 3 isomers ($E_{\gamma}\ge 1000\text{keV}$). Vertical axis $\times 10^{52}$}
        \label{fig:Figure4b}
    \end{minipage}

\setcounter{figure}{4}
\setcounter{subfigure}{-1}
\setlength{\abovecaptionskip}{+10pt}
    \caption{Evolution of the emitted $\gamma$-ray intensity, for the isomers relevant to GRBs, up to $t\le 3\text{s}$.}
    \label{fig:Figure4}
\end{subfigure}

%%%%%%%%% Figure 5 %%%%%%%%%%%%%%

\begin{figure}[h!]
\begin{center}
\includegraphics[width=1.0\textwidth]{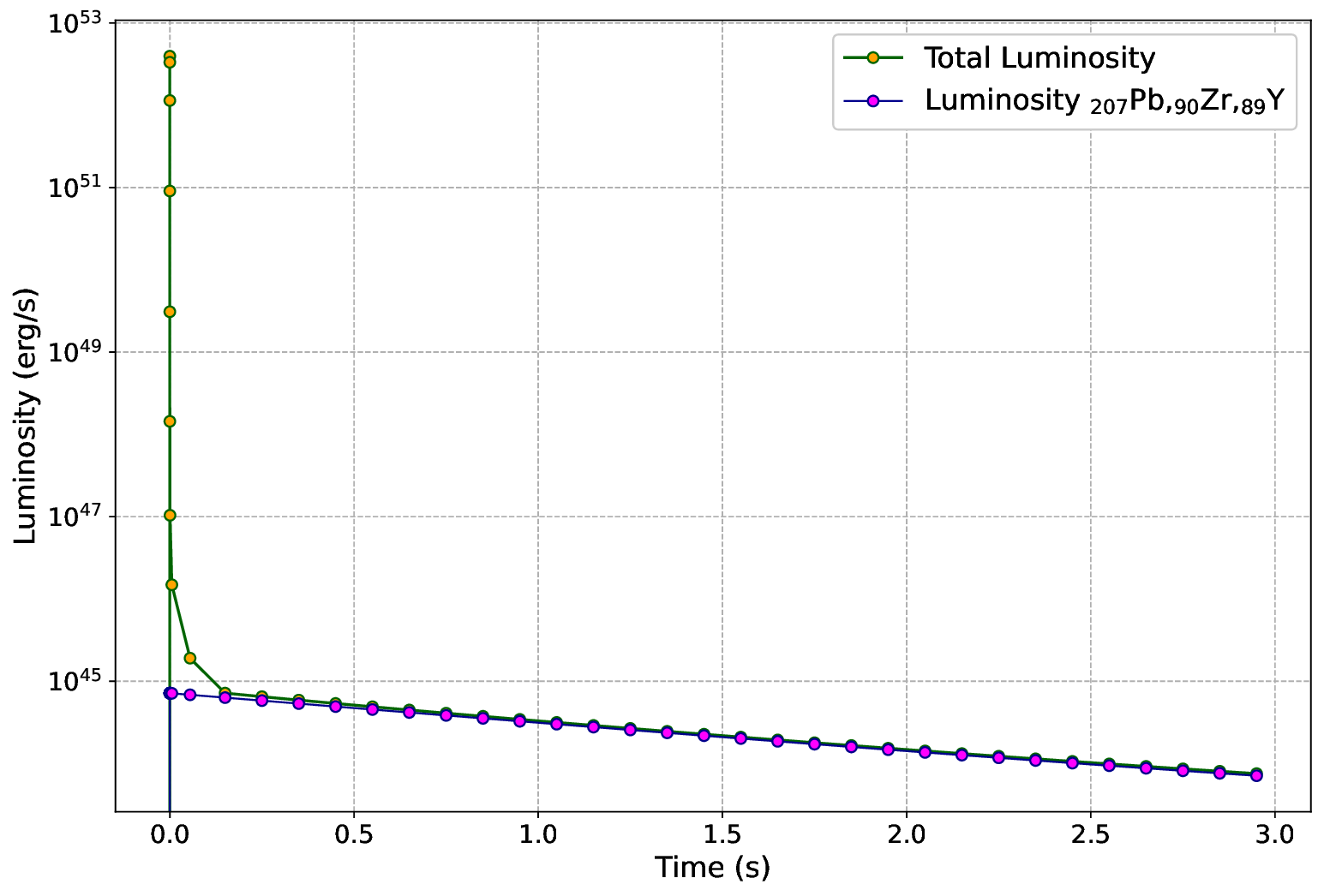}% This is a *.eps file
\end{center}
\caption{Comparison of total luminosity from all isomers against the
luminosity specifically from the three selected isomers, displayed on a logarithmic scale.}
\label{fig:Figure5}
\end{figure}

\end{document}